# An Analytical Framework for Understanding the Intensity of Religious Fundamentalism

**Navonil Bhattacharya**
LL.M, European Master in Law and Economics (EMLE)
e-mail: navonil.bhattacharya@gmail.com

**Arabinda Bhattacharya**
Former Associate Professor, Department of Business Management,
University of Calcutta, India
e-mail: ara_bha@yahoo.co.in

*Abstract*

*This paper examines the process of emergence of religious fundamentalism through development parameters. Therefore this research work reflects an analytical discussion on how the level of religious fundamentalism can be explained by the economic, political administrative and legal parameters such as GDP, Employment to Population ratio, Government Effectiveness, Voice & Accountability, Rule of Law (World Justice Project Report) and Rule of law (Governance Indicators).*

# 1. Introduction

The rise of religious fundamentalism can be viewed as a result of constant ideological conflict between the orthodox believer of religious verses and people who are not inclined to accept the religious dogmatism. However, it is not always true that only orthodox approach leads to religious fundamentalism. Religious fundamentalism is an outcome of series of interconnected socio-economic issues. Its inception might be religious conviction proclaimed by religious groups but when a section of the society starts acting as a part of counter cultural shock mechanism intentionally or otherwise, level of tolerance begins to tremble. This serious disagreement leads to violence and results in inevitable institutional failure. It is also true that the situation can be explained from the perspective of reverse causality. Institutional failure may aggravate the emergence of fundamentalism in other situation.

The emergence and aggravation of fundamentalism might not be only dependent on the propensity of conformist attitude of a certain group. Genesis of fundamentalism might also be dependent on the long history of ignorance, identity crisis and impoverishment. It is always easy to manipulate the ideology of a section of the society which is socially and economically vulnerable and uncertain about their legal rights.

The objective of the paper is to analyze whether development has an impact on the level of fundamentalism. The primary task of the paper is to define certain terms and concept relevant for this research and examine if there exists any relation between the development parameters and level of religious fundamentalism.

## 2. Definition of Religious Fundamentalism

Before plunging into the main discussion, certain concepts need to be revisited as these concepts will be appearing frequently in the analysis. The central proposition of the paper revolves around the religious fundamentalism and the factors contribute to its development. It is absolutely essential to present an explicit overview of fundamentalism for comprehending the importance of this paper. Any movement which employs coercion to limit the fundamental rights of a society or a particular group of the society would lead to fundamentalism. Usually, the term religious fundamentalism is used to denote an action of a group which is highly prejudiced by religious orthodoxy. Fundamentalist movement predominantly emerges from an urban society and disseminates a set of rules in regard to formation of societal structure, human behaviour and behaviour towards other. A set of contradictory principles vigorously encroaches the space of fundamental rights and strives to convert the essence of society from pluralism to monism. Almond, Sivan and Appleby (2003) defines fundamentalism as "a discernible pattern of religious militance by which self-styled 'true believers' attempt to arrest the erosion of religious identity, fortify the borders of the religious community, and create viable alternatives to secular institutions and behaviours." Religious Fundamentalist movements are viewed as "a religious response to modernity" (Paine, 1997). According to Paine (1997) - "Most scholars see the first major expression of modernity in the enlightenment, when thinkers rejected centralized and divinely sanctioned authority and instead relied on critical reason." Secular rationalists accept every fact with justified reason and every decision goes through a precise experimentation. This process are acknowledged universally as it is supposed to be superior to all other form of knowledge. On the other hand, fundamentalists reject the view of so-called "Modernity" as they only focus on traditionalist approach. In this approach, the people who are reactionaries act against the modern secular era. Fundamentalists believe that their existence is in a state of serious confusion due to identity crisis. This crisis leads to contradiction and a series of contradictions leads to conflicts. This maelstrom is intensified by the growing differentiation in the society. However, Andrew Paine (1997) anticipates a problem in a situation when the society and culture develop distinctively and they start to deviate from each other. If the alignment of the existing systems becomes dysfunctional, the society may experience such problems. Analyzing the literatures, it can be deduced that fundamentalism is a mixture of ideological and organizational variables which is again very dynamic in nature and the definition changes radically as certain cultural, economic, political and sociological premise change.

# 3. Survey of Literature

Andrew Paine has analysed the relationship between fundamentalists and legal system, and this relationship has been seen as an aspect of religious fundamentalism movement. Paine explained why the fundamentalists emphasize in dominating the legal system. Through this article, Paine explained the reason of growing attention on capturing the legal system by the fundamentalists and studied cases to prove the methods which fundamentalists followed to manipulate the system. Fundamentalists realizes the importance of dominating the legal system and decision making process to secure the future of society from counter-religious movements or secularism or modernism (Paine, 1997). Empirical studies prove that fundamentalists have been instrumental in impacting socio-economic factors, especially in defining "boundary maintenance" favourable for continuing their activity and to induce a section of a society as far as the political preferences (example- *presidential preference*) are concerned in American context (Hood and Morris, 1985). Martin E. Marty and R Scott Appleby are the pioneer scholars in the field of religious study. "*Fundamentalism Project*" is the most significant collaborative effort to provide a theoretical dimension of this complex concept till date. Historians, sociologists, cultural anthropologists have contributed to define, characterize and measure fundamentalism through their substantial research work. Marty, Appleby and other colleagues who were associated with this massive project have analyzed fundamentalism from social, political, cultural and theological perspectives. They have shed light on the concept of the emergence of fundamentalism through political and social theories. There was always a debate regarding the characterization of fundamentalist movement. This project identified nine (five ideological, four organizational) interrelated characteristics of fundamentalist groups (Almond, Sivan and Appleby, 1991). Laurence R. Iannaccone preferred the term "*Sectarian*" over fundamentalism while challenging all the previous assumptions regarding the characteristics and scope of fundamentalism and concluded that *Sectarianism* has a strong correlation with social & economic variables (Innaccone, 1997). Iannaccone argued that the people prefer to join sects because of indirect benefits which is induced by a cost-benefit analysis. There is always a debate among social and political theorists on measuring fundamentalism. Scholars are in serious disagreement while ascertaining the intensity of fundamentalism. The obvious question was what should be the parameters to determine if a movement or action can be termed as fundamentalism. Researchers are predominantly dependent on standardized characteristics (Almond, Sivan and Appleby, 1991). Religious movements need to be analyzed from the historical perspective in regard to categorize them as fundamentalist (Emerson and Hartman, 2006).

"*Religiopolitics*" or fundamentalism has been intensified across the globe due to unconstrained development in capitalism, disparity in income distribution, employment insecurities, forced migration, government favouritism towards a section of the society, and emergence of ideological and cultural clashes between migrants and original inhabitants (Keddie, 1998). Gang and Epstein (2007) have provided an economic model where rent seeking and asymmetric information problem have been identified as leading elements constituting fundamentalism. Fundamentalist leaders have the incentive to manipulate the members of the society in one hand and government on the other hand. Arce and Sandler (2003) used a game-theoretic model, specifically the Nash-demand game model to analyze the rise of fundamentalism. The research reveals that fundamentalists are more united and less compromising compared to others. It is very much evident from the literature review that the priority has been given to the theoretical aspect of religious fundamentalism for understanding the nature and elements constituting religious fundamentalist movements. Mainly, sociology, history and study of religion have been used as premise to examine the nature, development and future of religious fundamentalism. As far as the contribution of economic study is concerned, the volume of empirical studies are insubstantial compared to theoretical works. There is a clear distinction between the economic approach to examine religious fundamentalism and other social science studies. Social science has focused on the concept of modernity, impact of literal interpretation, threat on traditional values while postulating the emergence of religious fundamentalism. Whereas, economic approach has been concentrated on microeconomic theories of rent seeking, asymmetric information dilemma and cost benefit analysis. However, it is not completely true that researchers are not concerned about empirical studies related to religious fundamentalism. But empirical researches are restricted to selective fields, such as - sociology, psychology and religious study. But it is often ignored that the behaviour of a state, financial stability of a country and government's decision making process might lead to fundamentalist attitude and that should be analyzed from macroeconomic standpoint. Almond, Sivan and Appleby (*Explaining Fundamentalisms, 1995*) stated that Fundamentalists are predominantly enlisted from the less developed part of the society or from the rural population or from the poorer parts of urban areas, where people are less educated and the victim of asymmetrical economic and social development. This asymmetry may create intolerance and grievances among a particular section of the society in the population and tantalizes them into fundamentalist practices. Laurence Iannaccone (1997) has agreed on the fact that social and economic variables have immense importance in this upsurge of fundamentalism. The poor, less educated, and minority members of a society can be easily manipulated as they are ignored and discriminated by the dominating section of the society. Keddie (1998) has pointed out that income inequality, job insecurities and forced migration have immense impact on emergence of religious fundamentalism. According to Hood and Morris (1985), fundamentalists are capable of impacting political preferences. But a society in a terrible political chaos is also capable to give birth to a fundamentalist movement. People who are minority in a society are always marginalized from all aspects and for obvious reasons, they become extremely vulnerable towards fundamentalism. So, political preferences made by majority may lead to fundamentalist movements as well. The entire scenario becomes dependent upon government effectiveness, accountability of the individual and rule of law including multiple other macroeconomic parameters. Researchers have expressed their concern on socio-economic-political variables in theoretical discussion but it is hard to find comprehensive empirical researches which would shed some light on these variables. This paper might help to understand the issues of fundamentalism through macroeconomic modelling. This paper strives to find out the relevance of macroeconomic parameters in explaining the intensity of religious fundamentalism.

## 4. Research Objectives:

I. To find out the relevant socio-economic-legal Factors which will be characterising the country in question.
II. To examine whether there exists any functional relationship between the level of religious fundamentalism and the Factors stated above and also ascertain the strength of the relationship. In this context, the model which has been estimated is as follows:

$$LORF = f(F_1, F_2, F_3, \ldots)$$

Where, LORF: Level of Religious Fundamentalism, $F_i$: ith socio-economic-legal Factors

III. Given that there exists a functional relationship, to estimate the relative importance of the socio-economic-legal Factors in explaining the variations in the level of religious fundamentalism.
IV. To examine whether the estimation of the functional relationship changes across the different segmentations, segmentations being characterised by the level of development of the countries.

Therefore this research work reflects an analytical discussion on how the level of religious fundamentalism can be explained by the economic, political administrative and legal parameters such as GDP, Employment to Population ratio, Government Effectiveness, Voice & Accountability, Rule of Law etc..

## 5. Model

The following model is formulated so as to explain the nature of relationship between the level of fundamentalism and the development parameters. Theoretically, the development parameters are primarily categorized into three distinct groups, namely, Political & Administrative, Economic and Legal parameters which would play instrumental roles in determining the level of fundamentalism. It is conjectured that these three sets of parameters appear to have sufficient influence on the level of religious fundamentalism. Following the above logic, the model turns out to be as follows:

$$LORF = f(PA_1, PA_2, PA_3, ...., E_1, E_2, E_3, ...., L_1, L_2, L_3, ........)$$

Where, LORF: Level of Religious Fundamentalism, $PA_i$: ith Political & Administrative parameter, $E_i$: ith Economic parameter and $L_i$: ith Legal parameter.

If all these independent variables are related with one another, there might crop up the problem of multicollinearity for which it would be difficult, if not impossible, to estimate the above model and subsequently to disentangle the influence of each of the independent variables stated above. In order to get rid of this problem, the whole set of variables are mapped into different space ( may be with reduced no. of dimensions) where it is ensured that the new set variables so generated would be absolutely uncorrelated. These new set of variables are referred to as Factors and the analysis as Factor Analysis. Mathematically speaking, **($PA_1, PA_2, PA_3, ...., E_1, E_2, E_3, ...., L_1, L_2, L_3, ........$)** would be mapped with ($F_1, F_2, F_3, .......$) which will have one-to-one correspondence to the earlier set of variables. Since the new set of variables are strictly uncorrelated with one another, there will not be any problem of multicollinearity in estimating the following model:

$$LORF = f(F_1, F_2, F_3, ............)$$

More specifically, the model becomes as follows:

$$LORF = \beta_0 + \beta_1 F_1 + \beta_2 F_2 + \beta_3 F_3 + \beta_4 F_4 + ........... + \beta_n F_n + \epsilon$$

Where $\beta_0$: constant, $\beta_i$: coefficient associated with $F_i$ and $\epsilon$: error

In regression analysis, there is a measure like Multiple Correlation Coefficient ($R^2$), which would indicate how robust the model is. The standardized coefficients associated with above parameters would be indicative of their relative importance as far as the influence of these parameters on the level of religious fundamentalism is concerned.

## 6. Measuring Fundamentalism

Measuring fundamentalism is a difficult task as the measurement should be concerned with multiple variables reflecting the characteristics of fundamentalism. Therefore, any measure reflecting the level of fundamentalism is defined in a multidimensional space. In order to identify the degree of dominance of religious fundamentalism of each country considered for the analysis, A group of 15 EMLE (European Master in Law and Economics) students (considered to be an expert group) were requested to record their perception in regard to the intensity of fundamentalism that exists in each country. Initially 126 countries have been considered for this project. These countries have been by the above mentioned expert group based on their perception in regard to pervasiveness of fundamentalism of a particular country. These classification was done by a 7-point scale defined below.

7-point scale which is basically defined in an ordinal space would be defined in the following way:

1 : Absence of fundamentalism

2: Very low Level of fundamentalism

3: Somewhat low Level of fundamentalism

4: Moderately low Level of fundamentalism

5: Moderately high Level of fundamentalism

6: Sufficiently High Level of fundamentalism

7: Extreme Level of fundamentalism

Eventually, there would be multiple responses (no. of responses would be less than or equal to 15) corresponding to each country. Finally, each country has been classified on the basis of the average perception of the student. Median is taken as the average, the reason being that arithmetic mean cannot be justified with the ordinal data. Moreover, the median would be more appropriate to represent the central tendency of the data, because it minimises the biases at the both extreme ends of the distribution of multiple responses. The countries which could not be categorised in the 7-point scale by the experts, have been removed from the analysis. The survey has been conducted with 126 countries. However, due to the above reason, the number of countries has been reduced to 87 for final analysis.

## 7. Definition of Political & Administrative, Economic, and Legal Parameters

For this paper, it is absolutely necessary to prepare and define a set of development indicators which would consists of political & administrative, economic, legal indicators reflecting the overall performance of the country. As far as the Rule of Law is concerned, we have included two different Rule of Law variables. It is presumed that they might be able to capture all the relevant factors of legal performance.

**Table 1: Independent Variables**

| Independent Variables | Explanation | Abbreviation |
|---|---|---|
| **1. Political & Administrative Variables** | | |
| **1.1 Government Effectiveness: Estimate** | Government Effectiveness captures perceptions of the quality of public services, | **GOVTE** |

|  | the quality of the civil service and the degree of its independence from political pressures, the quality of policy formulation and implementation, and the credibility of the government's commitment to such policies. Estimate gives the country's score on the aggregate indicator, in units of a standard normal distribution, i.e. ranging from approximately - 2.5 to 2.5. |  |
|---|---|---|
| **1.2 Voice and Accountability: Estimate** | Voice and Accountability captures perceptions of the extent to which a country's citizens are able to participate in selecting their government, as well as freedom of expression, freedom of association, and a free media. Estimate gives the country's score on the aggregate indicator, in units of a standard normal distribution, i.e. ranging from approximately - 2.5 to 2.5. | **VAA** |
| *2. Economic Variables* | | |
| **2.1 GDP Growth (annual %)** | Annual percentage growth rate of GDP at market prices based on constant local currency. Aggregates are based on constant 2005 U.S. dollars. GDP is the sum of gross value added by all resident producers in the economy plus any product taxes and minus any subsidies not included in the value of the products. It is calculated without making deductions for depreciation of fabricated assets or for depletion and degradation | **GDPGR** |

| | of natural resources. | |
|---|---|---|
| **2.2 GDP per capita growth (annual %)** | Annual percentage growth rate of GDP per capita based on constant local currency. Aggregates are based on constant 2005 U.S. dollars. GDP per capita is gross domestic product divided by midyear population. GDP at purchaser's prices is the sum of gross value added by all resident producers in the economy plus any product taxes and minus any subsidies not included in the value of the products. It is calculated without making deductions for depreciation of fabricated assets or for depletion and degradation of natural resources. | **GDPPCGR** |
| **2.3 Employment to population ratio, 15+, total (%) (modelled ILO estimate)** | Employment to population ratio is the proportion of a country's population that is employed. Ages 15 and older are generally considered the working-age population. This variable is a part of Social Security and Labour index provided by World Bank. I have included this variable under the economic parameter group assuming the fact that the essence of this indicator has a linear relationship with the overall economic growth of a country. | **ETP** |
| **2.4 Income Group** | The countries are classified by Income Group, namely, High Income, Upper Middle Income, Lower Middle Income, and Low Income. This classification has been further classified into two i.e. High Income & Upper Middle Income as 1 | **IG** |

| | and Lower Middle Income & Low Income as 2 | |
|---|---|---|
| **3. Legal Variables** | | |
| **3.1 Rule of Law Index, World Justice Project Report, 2015** | This is a quantitative assessment tool which offers a detailed and comprehensive picture of the extent to which countries adhere to the rule of law in practice. Factors of the WJP Rule of Law Index include:<br>• Constraints on Government Powers<br>• Absence of Corruption<br>• Open Government<br>• Fundamental Rights<br>• Order and Security<br>• Regulatory Enforcement<br>• Civil Justice<br>• Criminal Justice | **ROLJR** |
| **3.2 Rule of Law: Estimate** | Rule of Law captures perceptions of the extent to which agents have confidence in and abide by the rules of society, and in particular the quality of contract enforcement, property rights, the police, and the courts, as well as the likelihood of crime and violence. Estimate gives the country's score on the aggregate indicator, in units of a standard normal distribution, i.e. ranging from approximately - 2.5 to 2.5. | **ROLGI** |

*Source: World Bank Governance Indicators, 2014, World Bank Rule of Law Index, 2014 and World Justice Project, 2015*

# 8. Analysis 1: Factor Analysis and Regression Analysis

The following steps are followed for analysing the data set.

*Step 1*: Formation of Factors

Factor analysis has been done with six variables, namely - ROLJR, ROLGI, GOVTE, VAA, ETP and GDPGR in order to eliminate the problem of multicollinearity in the estimation of the regression model.

**Table 2: Factor Analysis - KMO and Bartlett's Test**

**KMO and Bartlett's Test**

| | | |
|---|---|---|
| Kaiser-Meyer-Olkin Measure of Sampling Adequacy. | | .793 |
| Bartlett's Test of Sphericity | Approx. Chi-Square | 608.726 |
| | df | 15 |
| | Sig. | .000 |

Results in Table 2 ensure the justification of factor analysis. High value of KMO measure of sampling adequacy (0.793) and also the very low value of significance indicate that the variables considered for the analysis are strongly correlated to each other.

**Table 3: Factor Analysis - Total Variance Explained**

**Total Variance Explained**

| Component | Initial Eigenvalues | | | Extraction Sums of Squared Loadings | | |
|---|---|---|---|---|---|---|
| | Total | % of Variance | Cumulative % | Total | % of Variance | Cumulative % |
| 1 | 3.808 | 63.467 | 63.467 | 3.808 | 63.467 | 63.467 |
| 2 | 1.289 | 21.483 | 84.950 | 1.289 | 21.483 | 84.950 |
| 3 | .564 | 9.402 | 94.352 | | | |
| 4 | .260 | 4.341 | 98.693 | | | |
| 5 | 5.132E-02 | .855 | 99.548 | | | |
| 6 | 2.712E-02 | .452 | 100.000 | | | |

Extraction Method: Principal Component Analysis.

In the above table 3, it is shown that the first two factors have eigenvalues less than 1 and these two factors explains significant portion of the total variations i.e. 84.95%. Therefore, only first two factors are considered for the further analysis.

**Table 4: Factor Analysis - Component Matrix**

**Component Matrix<sup>a</sup>**

|  | Component 1 | Component 2 |
|---|---|---|
| ROLJR | .971 | .137 |
| ROLGI | .962 | .197 |
| GOVTE | .945 | .146 |
| VAA | .898 | -6.01E-03 |
| ETP | -.158 | .859 |
| GDPGR | -.464 | .687 |

Extraction Method: Principal Component Analysis.
a. 2 components extracted.

From the component matrix, it is clear that the first factor has very strong correlation with ROLJR, ROLGI, GOVTE and VAA. Therefore the first factor is named as Legal & Administrative variable (LAV). Similarly, two economic parameters, such as, ETP and GDPGR are reflected in second factor. Hence, it is named as Economic Variable (EV). Hence the analysis leads to formation of two factors -

**Factor1**: Legal and Political & Administrative variables ( LAV ) is considered to be reflection of the parameters, namely, ROLJR, ROLGI, GOVTE, and VAA.

**Factor2**: Economic variables (EV) is concerned with GDPGR and ETP

*Step 2*: **Estimation of Regression Model**

**Table 5: Value of $R^2$**

**Model Summary**

| Model | R | R Square | Adjusted R Square | Std. Error of the Estimate |
|---|---|---|---|---|
| 1 | .565[a] | .319 | .303 | 1.3093 |

a. Predictors: (Constant), FACTOR2:EV FACTOR1: LAV

It is clear from the table 5 that Model is fairly robust, because Economic Variable (EV) and Legal & Administrative Variable (LAV) explains 31.9% of variation of the dependent variable, i.e. Level of Religious Fundamentalism (LORF). Robustness of the Model is further established in the following table 6 by very small value of significance.

**Table 6: ANOVA for the Regression Model**

**ANOVA[b]**

| Model | | Sum of Squares | df | Mean Square | F | Sig. |
|---|---|---|---|---|---|---|
| 1 | Regression | 67.550 | 2 | 33.775 | 19.701 | .000[a] |
| | Residual | 144.008 | 84 | 1.714 | | |
| | Total | 211.557 | 86 | | | |

a. Predictors: (Constant), FACTOR2:Economic Factor, FACTOR1: Legal & Administrative Factor

b. Dependent Variable: LORF

**Table 7: Regression Results**

**Coefficients<sup>a</sup>**

| Model | | Unstandardized Coefficients | | Standardized Coefficients | t | Sig. |
|---|---|---|---|---|---|---|
| | | B | Std. Error | Beta | | |
| 1 | (Constant) | 2.862 | .141 | | 20.329 | .000 |
| | FACTOR1: Legal & Administrative Factor | -.860 | .143 | -.543 | -6.027 | .000 |
| | FACTOR2: Economic Factor | -.231 | .148 | -.140 | -1.558 | .123 |

a. Dependent Variable: LORF

From the regression results, it can be concluded that LAV i.e. Legal & Administrative variable has turned out to be statistically highly significant. This means that LAV has significant impact on the level of Religious Fundamentalism and the nature of relationship is of inverse type. To be precise, if LAV increases, the LORF is likely to decrease. This result can be interpreted in the following way. If the rule of law and also the administrative aspect are stringent, with all probability the level of religious fundamentalism is bound to be reduced leading to conducive environment for the community. This is also interesting to note that the economic parameters do not appear to be at all important in determining the level of religious fundamentalism at the macro level.

## 9. Analysis 2: Income Group-wise Regression Analysis

In this section, the countries considered for the analysis are segmented on the basis of the level of socio-economic-legal parameters published by The world Justice project (WSP) Rule of Law Index 2016. The basis for this segmentation is taken as income group the countries. According to the report the countries are categorised into 4, namely, High Income, Upper Middle Income, Lower Middle Income and finally Low Income. In this research work, the data set is broadly divided into two groups, namely, High Income & Upper Middle Income ( Group 1 ) and Lower Middle Income & Low Income ( Group 2). Separate analysis has been undertaken for these two groups.

## 9.1 Results for High Income & Upper Middle Income ( Group 1 )

**Table 8: Value of $R^2$**

**Model Summary**

| Model | R GROUP = high & upper middle income (Selected) | R Square | Adjusted R Square | Std. Error of the Estimate |
|---|---|---|---|---|
| 1 | .512[a] | .262 | .235 | 1.1864 |

a. Predictors: (Constant), FACTOR2:Economic Factor, FACTOR1: Legal & Administrative Factor

**Table 9: ANOVA for the Regression Model**

**ANOVA[b,c]**

| Model | | Sum of Squares | df | Mean Square | F | Sig. |
|---|---|---|---|---|---|---|
| 1 | Regression | 27.485 | 2 | 13.743 | 9.764 | .000[a] |
| | Residual | 77.411 | 55 | 1.407 | | |
| | Total | 104.897 | 57 | | | |

a. Predictors: (Constant), FACTOR2:Economic Factor, FACTOR1: Legal & Administrative Factor

b. Dependent Variable: LORF

c. Selecting only cases for which GROUP = high & upper middle income

From the table 8 and 9, it is clear that the independent variables like EV and LAV explain the significant portion of the dependent variable i.e. LORF and value of $R^2$ turns out to be 0.262 which is statistically significant reflected in table 9.

**Table 10: Regression Results for the Higher Income Group**

**Coefficients**[a,b]

| Model | | Unstandardized Coefficients | | Standardized Coefficients | t | Sig. |
|---|---|---|---|---|---|---|
| | | B | Std. Error | Beta | | |
| 1 | (Constant) | 2.798 | .195 | | 14.356 | .000 |
| | FACTOR1: Legal & Administrative Factor | -.830 | .198 | -.529 | -4.186 | .000 |
| | FACTOR2: Economic Factor | 7.384E-02 | .192 | .049 | .384 | .702 |

a. Dependent Variable: LORF
b. Selecting only cases for which GROUP = high & upper middle income

Like the overall regression model, the specific model for the Higher Income group indicate that in determining the level of fundamentalism, it is LAV (Legal & Administrative Variable) which becomes very important. It does have significant influence on the level of religious fundamentalism in the negative way. EV(Economic Variable) does seem to be insignificant.

## 9.2 Results for the Lower Middle Income & Low Income ( Group 2)

**Table 11: Value of $R^2$**

**Model Summary**

| Model | R GROUP = low & lower middle income (Selected) | R Square | Adjusted R Square | Std. Error of the Estimate |
|---|---|---|---|---|
| 1 | .535[a] | .287 | .232 | 1.3936 |

a. Predictors: (Constant), FACTOR2:Economic Factor, FACTOR1: Legal & Administrative Factor

## Table 12: ANOVA for the Regression Model

**ANOVA[b,c]**

| Model | | Sum of Squares | df | Mean Square | F | Sig. |
|---|---|---|---|---|---|---|
| 1 | Regression | 20.296 | 2 | 10.148 | 5.225 | .012[a] |
| | Residual | 50.497 | 26 | 1.942 | | |
| | Total | 70.793 | 28 | | | |

a. Predictors: (Constant), FACTOR2:Economic Factor, FACTOR1: Legal & Administrative Factor

b. Dependent Variable: LORF

c. Selecting only cases for which GROUP = low & lower middle income

Table 11 and 12 show that LAV and EV together explain 28.7% of the total variation of the level of Fundamentalism. Table 12 ensures the level of significance being at the order of 1.2%

## Table 13: Regression Results for the Lower Income Group

**Coefficients[a,b]**

| Model | | Unstandardized Coefficients | | Standardized Coefficients | t | Sig. |
|---|---|---|---|---|---|---|
| | | B | Std. Error | Beta | | |
| 1 | (Constant) | 2.893 | .536 | | 5.395 | .000 |
| | FACTOR1: Legal & Administrative Factor | -1.215 | .586 | -.370 | -2.073 | .048 |
| | FACTOR2:Economic Factor | -.848 | .275 | -.550 | -3.077 | .005 |

a. Dependent Variable: LORF

b. Selecting only cases for which GROUP = low & lower middle income

This is very interesting to note that for the Lower Income Group, both the variables i.e. Legal & Administrative variable (LAV) and Economic variable (EV) turn out to be statistically significant. Both the independent variables are negatively related to the level of fundamentalism. So this can be concluded from the regression results that the level of Fundamentalism can be checked and restricted to a tolerable limit, if stringent legal & administrative measures and also the economic development at relatively higher level are ensured in the country. This is also clear from the standardized coefficients that of the two variables, Economic variable seems to be more sensitive in determining the Level of Religious Fundamentalism.

Therefore, it is absolutely important for the countries belonging to the lower income group to take up projects ensuring economic development along with effective legal and administrative measures in order to reduce the level of fundamentalism to a significant extent.

## 10. Conclusion

The growing incidence of religious fundamentalism across the globe is a serious challenge for the entire mankind. Studies on this issue have shown an inclination towards the concept of 'Modernism' in the context of pivotal elements which are instrumental for the rise of religious fundamentalism. There are also massive literatures which have made theoretical attempt to explore other economic factors responsible for the rise of religious fundamentalism. This paper examines the importance of those economic, social, political, administrative and legal parameters and it provides a stochastic estimate to discern the intensity of fundamentalism. Macro-level determinants have been used to shed light on the role of development parameters *vis-à-vis* the intensity of fundamentalism. The study might help to prioritize the critical aspects of the issue setting aside the emergence of significant consequences of fundamentalism. The essence of research indicates how the government effectiveness, voice and accountability, economic factors, rule of law and so on might be instrumental in determining the level of fundamentalism. Between two factors, namely, Legal & Administrative factor (LAV) and Economic factor(EV), legal & administrative factor has turned out to be decisive factor which should concern the policymakers and overall legal governance system at the macro level.

Empirical evidence from lower income group countries suggests that both legal and economic factors have controlling effect on fundamentalism, whereas, in the case of higher income group countries, only the legal and administrative factor has come out to be significant. An obvious dissimilarity between the results of two income groups might stipulate the respective government to formulate efficient policies. It is evident from the study that social, legal and economic vulnerabilities are more likely to end up in influencing the fundamentalist attitude. It can be interpreted with the help of empirical realities that political economy of religious fundamentalism has the potential to dictate human behaviour who are marginalized by multi dimensional development parameters. Inflexible and biased attitude of the state towards marginalized section might lead them to be engaged in more pro-fundamentalist activities as the legal, political, administrative and economic spaces are being reduced.

Governments should start revamping the policies and procedures in regard to the vulnerable section of the society. Legislators and policymakers must concentrate on (a) setting the economic priority for the vulnerable section of the society (b) strengthening the voice and accountability mechanism (c) enhancing the effectiveness of the government and reducing the biasness (d) empowering marginalized section to have legal recourse. In a nutshell, it can be concluded that in this extremely critical process of eliminating the religious fundamentalism, states should concentrate more on the root cause analysis rather than debating on the consequences.

## References

- Almond, G.A., Appleby, R.S., and Sivan, E. (2003). Strong religion: The rise of fundamentalisms around the world. University of Chicago Press: Chicago and London.

- Arce, D. G. and Sandler, T. (2003). An evolutionary game approach to fundamentalism and conflict. Journal of Institutional and Theoretical Economics, 159 (1), 132-154.

- Arce, D. G. and Sandler, T. (2009). Fitting in: group effects and the evolution of fundamentalism. Journal of Policy Modelling,31 (5), 739-757.

- Berman, E. (2000). Sect, subsidy and sacrifice: an economist's view of ultra-orthodox Jews. Quarterly Journal of Economics, 115 (3), 905-953.

- Emerson, M.O., and Hartman, D. (2006). The rise of religious fundamentalism. Annual Review of Sociology, 32, 127-144.

- Epstein, G.S., and Gang, I. (2007). Understanding the development of fundamentalism. Public Choice, 132 (3-4), 257-271.

- Hood, R.W., Hill, C.H., and Williamson, W.P. (2005). The psychology of religious fundamentalism. The Guildford Press: New York and London.

- Hood, R.W., and Morris, R. (1985). Boundary Maintenance, Social-Political Views. and Presidential Preference among High and Low Fundamentalists, Review of Religious Research, Vol. 27, No. 2.

- Iannaccone, L.R. (1992). Sacrifice and stigma: reducing free-riding in cults, communes and other collectives. Journal of Political Economy, 100, 271-92.

- Iannaccone, L.R. (1997). Towards an economic theory of `fundamentalism.' Journal of Institutional and Theoretical Economics, 153 (1), 100-116.

- Keddie, N. R. (1998). The new religious politics: where, when, and why do 'fundamentalisms' appear? Comparative Studies in Society and History, 40, 696-723.

- Levy, G., and Razin, R. (2012). Religious beliefs, religious participation, and cooperation. American Economic Journal: Microeconomics, 4 (3), 121-151.

- Lord, Charles G., Lee Ross, and Mark R. Lepper (1979). Biased assimilation and attitude polarization: The effects of prior theories on subsequently considered evidence. Journal of personality and social psychology 37 (11), 2098-2109.

- Makowsky, M. D. (2012), Emergent extremism in a multi-agent model of religious clubs. Economic Inquiry, 50, 327-347.

- Marty, M.E., and Appleby, R.S. (Eds.). (1991). Fundamentalism project: fundamentalisms observed, Vol.1. University of Chicago Press: Chicago.

- McBride, M. (2015). Why churches need free-riders: Religious capital formation and religious group survival. Journal of Behavioral and Experimental Economics, 58(C),77-87.

- Paine, A. (1997). Religious Fundamentalism and Legal Systems: Methods and Rationales in the Fight to Control the Political Apparatus, Indiana Journal of Global Legal Studies, Vol-1, Issue 5.

- Sen, A. (2008). Violence, Identity and Poverty, Journal of Peace Research, Vol. 45, PP. 5-15

- Shy, O. (2007). Dynamic models of religious conformity and conversion: Theory and calibrations. European Economic Review, 51 (5), 1127-1153.